\documentclass[amsmath,amssymb,superscriptaddress,nobalancelastpage,aps,twocolumn,showpacs]{revtex4-1}

\usepackage{array,multirow,graphicx}
\usepackage{braket}
\usepackage{epstopdf}
\usepackage{graphicx}
\usepackage{siunitx}
\usepackage{nicefrac}
\usepackage[normalem]{ulem}
\usepackage{varioref}
\usepackage{xcolor}
\usepackage{xfrac}
\usepackage{xr-hyper}
\usepackage{hyperref}

\usepackage{physics}

\definecolor{bl}{rgb}{0.0,0.2,0.6}

\hypersetup{colorlinks,linkcolor=blue,urlcolor=blue,citecolor=blue}

\newcommand{\PLCCO}{Pr$_{1.3-x}$La$_{0.7}$Ce$_{x}$CuO$_{4}$}
\newcommand{\RCCO}{{\it R}$_{2-x}$Ce$_{x}$CuO$_{4}$}
\newcommand{\LCCO}{La$_{2-x}$Ce$_{x}$CuO$_{4}$}
\newcommand{\PCCO}{Pr$_{2-x}$Ce$_{x}$CuO$_{4}$}
\newcommand{\NCCO}{Nd$_{2-x}$Ce$_{x}$CuO$_{4}$}

\begin{document}
  \author{M.~Miyamoto}
	\affiliation{Institute for Solid State Physics, The University of Tokyo, Kashiwa, Chiba 277-8581, Japan}

	\author{M.~Horio}
	\email{mhorio@issp.u-tokyo.ac.jp}
	\affiliation{Institute for Solid State Physics, The University of Tokyo, Kashiwa, Chiba 277-8581, Japan}
	
	\author{K.~Moriya}
	\affiliation{Department of Engineering and Applied Sciences, Sophia University, 7-1 Kioi-cho, Chiyoda-ku, Tokyo 102-8554 Japan}

	\author{A.~Takahashi}
	\affiliation{Department of Applied Physics, Tohoku University, 6-6-05 Aoba, Aramaki, Aoba-ku, Sendai 980-8579, Japan}
	
	\author{J.~Osiecki}
	\affiliation{MAX IV Laboratory, Lund University, Lund SE-221 00, Sweden}

	\author{B.~Thiagarajan}
	\affiliation{MAX IV Laboratory, Lund University, Lund SE-221 00, Sweden}

	\author{C.~M.~Polley}
	\affiliation{MAX IV Laboratory, Lund University, Lund SE-221 00, Sweden}

	\author{Y.~Koike}
	\affiliation{Department of Applied Physics, Tohoku University, 6-6-05 Aoba, Aramaki, Aoba-ku, Sendai 980-8579, Japan}

	\author{T.~Adachi}
	\affiliation{Department of Engineering and Applied Sciences, Sophia University, 7-1 Kioi-cho, Chiyoda-ku, Tokyo 102-8554 Japan}

	\author{T.~Mizokawa}
	\affiliation{Department of Applied Physics, Waseda University, Shinjuku, Tokyo 169-8555, Japan}

	\author{I.~Matsuda}
	\affiliation{Institute for Solid State Physics, The University of Tokyo, Kashiwa, Chiba 277-8581, Japan}
	
	\title{Inhomogeneous reduction-annealing effects on the electron-doped cuprate superconductor revealed by micro-focused angle-resolved photoemission spectroscopy}
	
\begin{abstract}
	\indent The development of the protect-annealing method has extended the superconductivity of the 
	electron-doped cuprate Pr$_{1.3-x}$La$_{0.7}$Ce$_{x}$CuO$_{4}$ (PLCCO) into lower Ce concentrations, while the superconducting volume 
	fraction decreases with underdoping. Employing angle-resolved photoemission spectroscopy with a micro-focused 
	beam, we investigated the electronic structure of protect-annealed PLCCO ($x=0.08$) with small 
	superconducting volume fraction. Significant spatial variation of Fermi surface area and shape was 
	observed, suggesting inhomogeneity in electron concentrations and the pseudogap that competes with 
	superconductivity. By performing measurements at dozens of different sample positions, negative and 
	non-monotonic correlation was found between the electron concentration and pseudogap magnitude. 
	The established correlation illustrates a systematic annealing dependence of the electronic structure 
	where a pseudogap abruptly opens with insufficient oxygen reduction. 
\end{abstract}
	
\maketitle
\section{Introduction}
 For electron-doped cuprate superconductors \RCCO ~({\it R}: rare earth), it has been widely 
accepted that post-growth reduction annealing is essential for the emergence of superconductivity 
\cite{takagi1989superconductivity,tokura1989superconducting}. Irrespective of the amount of Ce 
substitutions for electron doping, superconductivity is induced only after reduction annealing. 
It was found by neutron-scattering studies that as-grown samples contain impurity oxygen atoms 
at the apical site, right above the Cu site, whose occupation decreases after annealing 
\cite{radaelli1994evidence,schultz1996single,fujita2021reduction}. Applying stronger reducing conditions, superconducting 
transition temperature $T_{\rm c}$ is enhanced likely owing to more efficient removal of apical oxygen 
atoms \cite{kim1993phase}. In this context, thin-film samples have an advantage of large surface-to-volume ratio that 
should facilitate oxygen diffusion out of the sample. Indeed, it was demonstrated that superconductivity can be realized 
in thin films at higher temperatures and, surprisingly, even without Ce substitutions \cite{tsukada2005role,matsumoto2009generic}, 
though the parent compound {\it R}$_{2}$CuO$_{4}$ has long been believed to be a Mott insulator \cite{armitage2010progress}. 
This discovery has stimulated extensive studies to elucidate the essential ingredients of superconductivity 
in electron-doped cuprates as well as the influence of reduction annealing on their electronic structure.\\
\indent In order to address the issue using bulk single crystals, Adachi {\it et al.} 
\cite{adachi2013evolution} developed the so-called protect-annealing method for more efficient 
reduction. With the technique, they realized superconductivity in \PLCCO ~(PLCCO) at $x=0.10$, 
which was reported to be insulating in previous studies under conventional annealing \cite{sun2004thermal}. 
An angle-resolved photoemission spectroscopy (ARPES) study of protect-annealed PLCCO ($x=0.10$) \cite{horio2016suppression} 
revealed dramatic suppression of a pseudogap that competes with superconductivity, while the electron 
concentration estimated from Fermi surface area was significantly larger than the Ce concentration. 
The latter observation suggests that not only oxygen atoms at the apical site but also those at the 
regular sites (CuO$_2$ planes and {\it R}$_2$O$_2$ layers) were removed to increase the electron concentrations 
in the CuO$_2$ plane \cite{horio2018electronic,horio2018angle,lin2021extended}. It thus remains elusive if the suppression 
of the pseudogap and emergence of superconductivity in the Ce-underdoped sample were driven by the removal of impurity 
oxygen atoms or by the increase of electron concentrations. Upon further decreasing Ce concentrations of bulk 
single crystals, superconductivity is still attained but only with small volume fraction. For such a 
crystal of protect-annealed PLCCO ($x=0.05$), Matsuzawa {\it et al.} \cite{matsuzawa2021fermi} performed 
angle-integrated photoemission spectroscopy measurements using a nano-focused beam and found that the 
spectral intensity of the charge-transfer band, which is sensitive to electron doping \cite{armitage2002doping}, 
exhibits microscopic spatial variation probably due to unbalanced oxygen diffusion. In order to capture 
the influence of protect-annealing on Ce-underdoped samples properly, such spatial variation should be taken into account.\\
\indent Here, we present an ARPES study of protect-annealed PLCCO ($x=0.08$) bulk single crystals with 
small superconducting volume fraction using a microbeam focused to 10 ${\rm\mu}$m$\times$10 ${\rm\mu}$m. The area 
and shape of the observed Fermi surface significantly varied over space, suggesting inhomogeneous electronic 
states. By evaluating electron concentrations and pseudogap magnitude at dozens of different sample 
positions, we established a correlation between both quantities from a single sample. The obtained 
relationship presents a systematic annealing dependence of the electronic structure where a pseudogap 
abruptly opens with insufficient oxygen reduction, which suggests the major influence of electron concentrations 
as well as significant effect of apical oxygen atoms.\\

\section{Methods}
Single crystals of PLCCO ($x=0.08$) were synthesized by the traveling-solvent floating-zone 
method \cite{adachi2013evolution}, and as-grown samples showed no superconductivity. Post-growth reduction 
annealing was conducted using the protect-annealing method at 800 ℃ for 24 hours. 
\begin{figure}
  \begin{center}
    \includegraphics[width=0.45\textwidth]{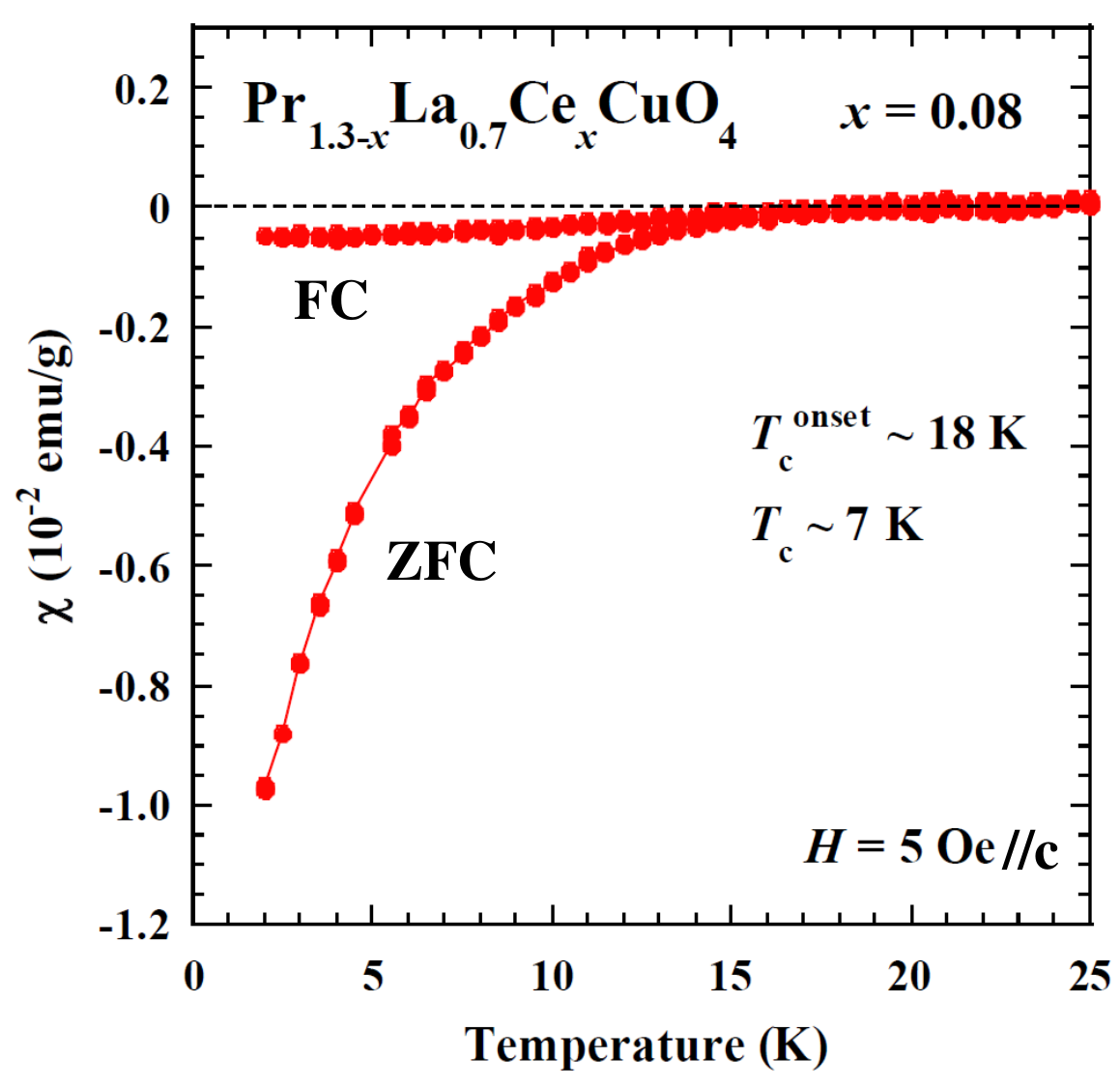}
  \end{center}
  \caption{\textbf{Superconductivity in PLCCO with small superconducting volume fraction.} 
	Magnetic susceptibility of a protect-annealed PLCCO ($x=0.08$) sample. The measurement was 
	performed with field cooling (FC) and zero-field cooling (ZFC). The onset temperature of 
	superconducting transition is $T^{\rm{onset}}_{\rm{c}}\sim$18 K, and the transition 
	temperature obtained by extrapolating the steepest slope is $T_{\rm{c}}\sim$7 K.
	The magnetic field of 5 Oe was applied along the {\it c} axis.
	}
  \label{fig.1}
\end{figure}
Figure~\ref{fig.1} shows magnetic susceptibility of the protect-annealed single crystal 
of PLCCO ($x=0.08$). The transition temperature $T_{\rm c}$ obtained by extrapolating 
the steepest slope is $\sim$7 K. The diamagnetic signal is comparable to that of 
the previously studied PLCCO ($x=0.05$) sample and is 3-4 times smaller than the typical value for 
the protect-annealed $x=0.10$ samples with similar dimensions \cite{matsuzawa2021fermi}.
This suggests that the present crystal has relatively small superconducting volume fraction.\\
\indent ARPES measurements were performed for both the as-grown and annealed samples 
at the BLOCH beamline of MAX IV. The incident beam was focused to 10 ${\rm\mu}$m$\times$ 10 ${\rm\mu}$m. 
Using the micro-focused beam, we collected data at various positions within the relatively 
flat surface of $\sim$800 ${\rm\mu}$m$\times$ $\sim$600 ${\rm\mu}$m for every sample. 
Incident photon energy was set at 55 eV and total energy resolution at 10 meV. The samples were 
cleaved {\it in-situ} and measured at $T=20$ K under the vacuum better than $1\times10^{-10}$ mbar. 

\section{Results}
\begin{figure}
  \begin{center}
    \includegraphics[width=0.5\textwidth]{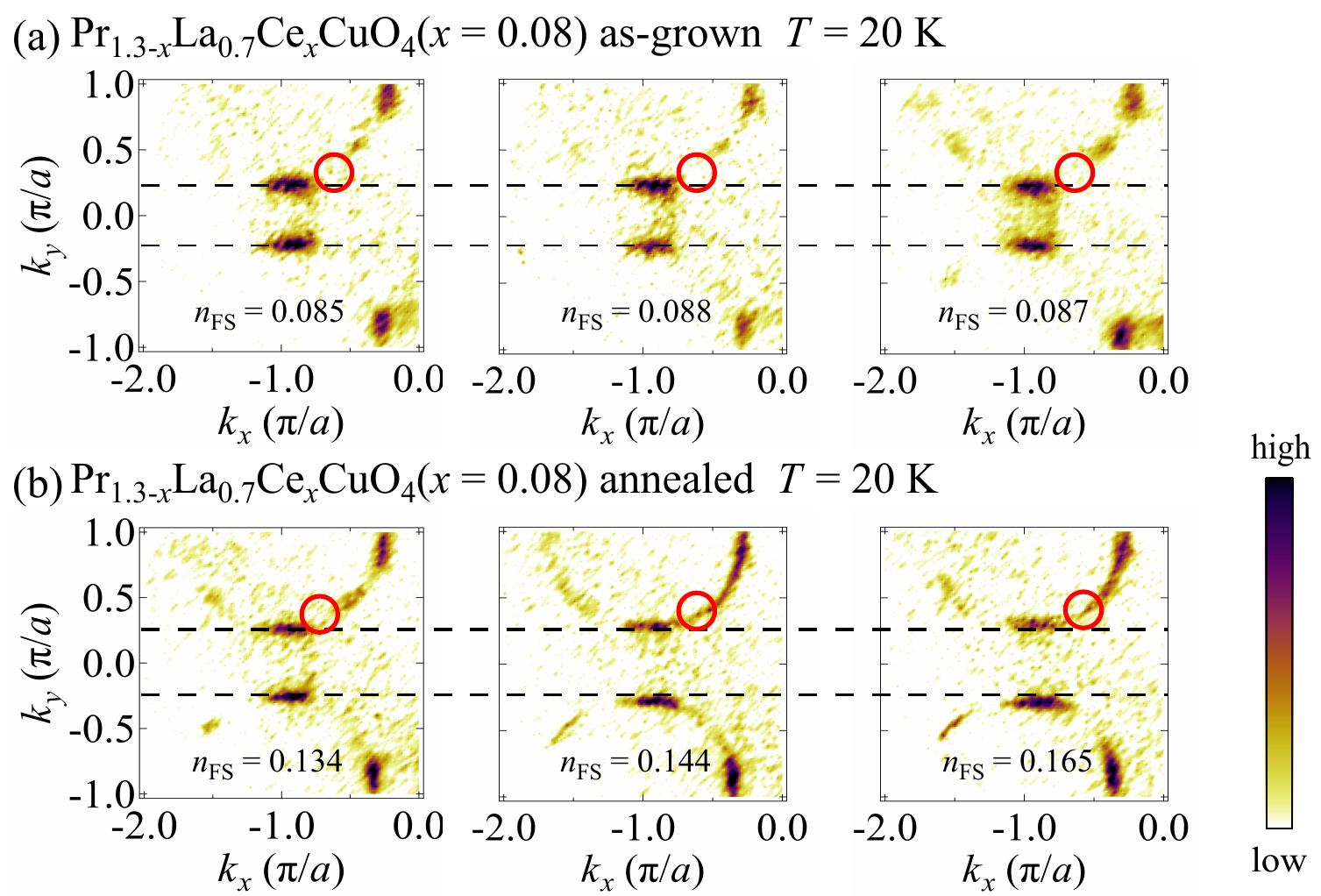}
  \end{center}
  \caption{\textbf{Inhomogeneous electronic states.} 
	FS maps of PLCCO ($x=0.08$) at different positions are shown for 
	(a) the as-grown sample and for (b) the protect-annealed sample. 
	The spectral intensity is integrated over $\pm 20$ meV around the Fermi level $E_{\rm{F}}$. 
	The black dashed lines represent the distance between antinodal segments in the leftmost panel 
	for each sample. The as-grown sample has almost the same antinodal distance and electron concentration 
	$n_{\rm FS}$, as well as comparable intensity at the hot spot marked with the red circle. In contrast, 
	the annealed sample shows significant position dependence. In going from the left to right panels, 
	the distance between antinodal parts expands, the electron concentration increases, and the intensity at 
	the hot spot becomes stronger.
	}
  \label{fig.2}
\end{figure}
\begin{figure}
  \begin{center}
    \includegraphics[width=0.5\textwidth]{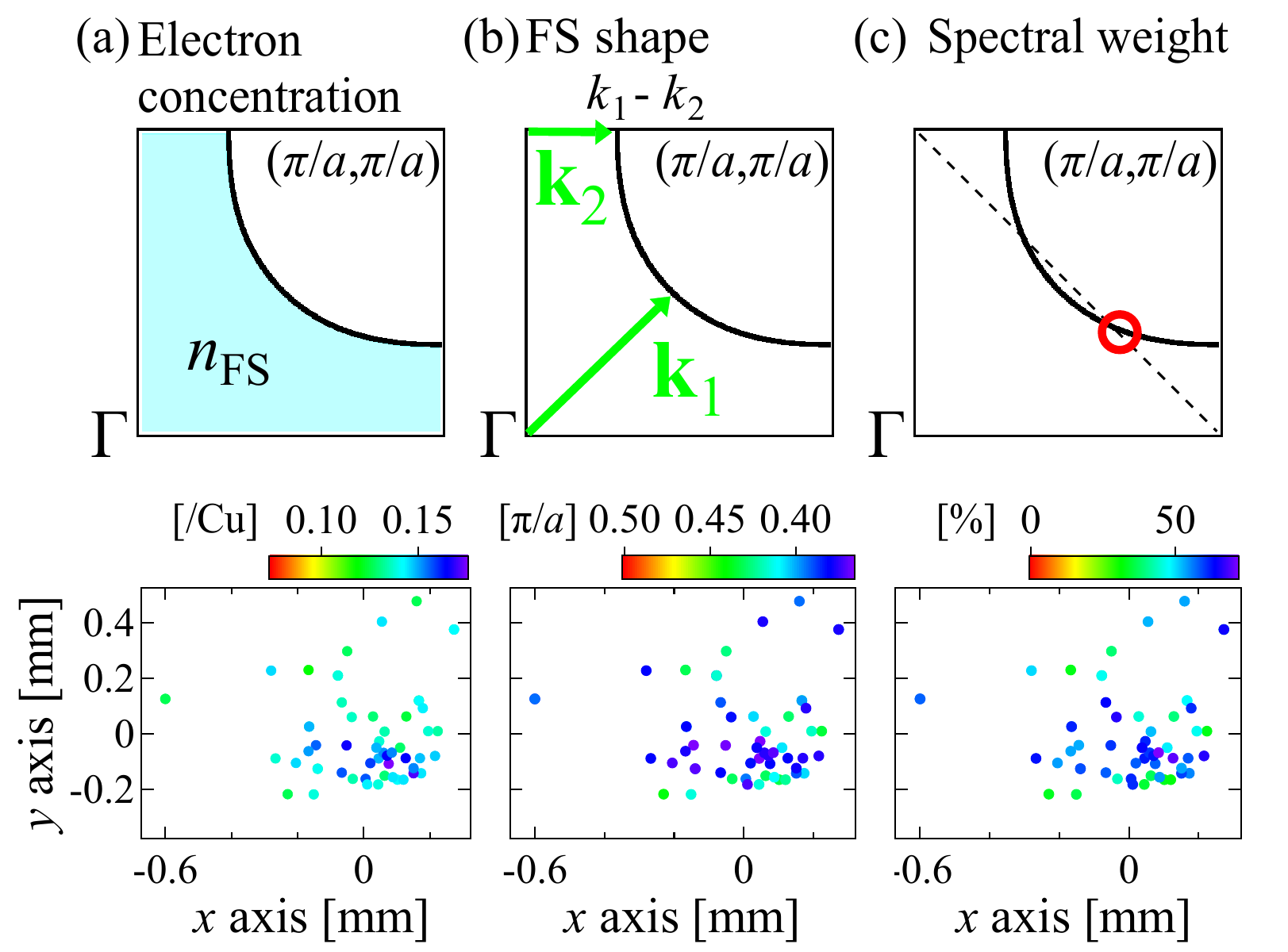}
  \end{center}
  \caption{\textbf{Spatial variation of physical quantities evaluated for the protect-annealed sample.} 
	Measurement-position dependence of (a) the electron concentration $n_{\rm{FS}}$ 
	determined from Fermi surface area, (b) FS shape evaluated from the two	wave vectors 
	$\vb{k_1}$ and $\vb{k_2}$, and (c) spectral weight at the hot spot 
	integrated over $\pm50$ meV around $E_{\rm F}$.
	}
  \label{fig.3}
\end{figure}
\begin{figure*}
  \begin{center}
    \includegraphics[width=0.9\textwidth]{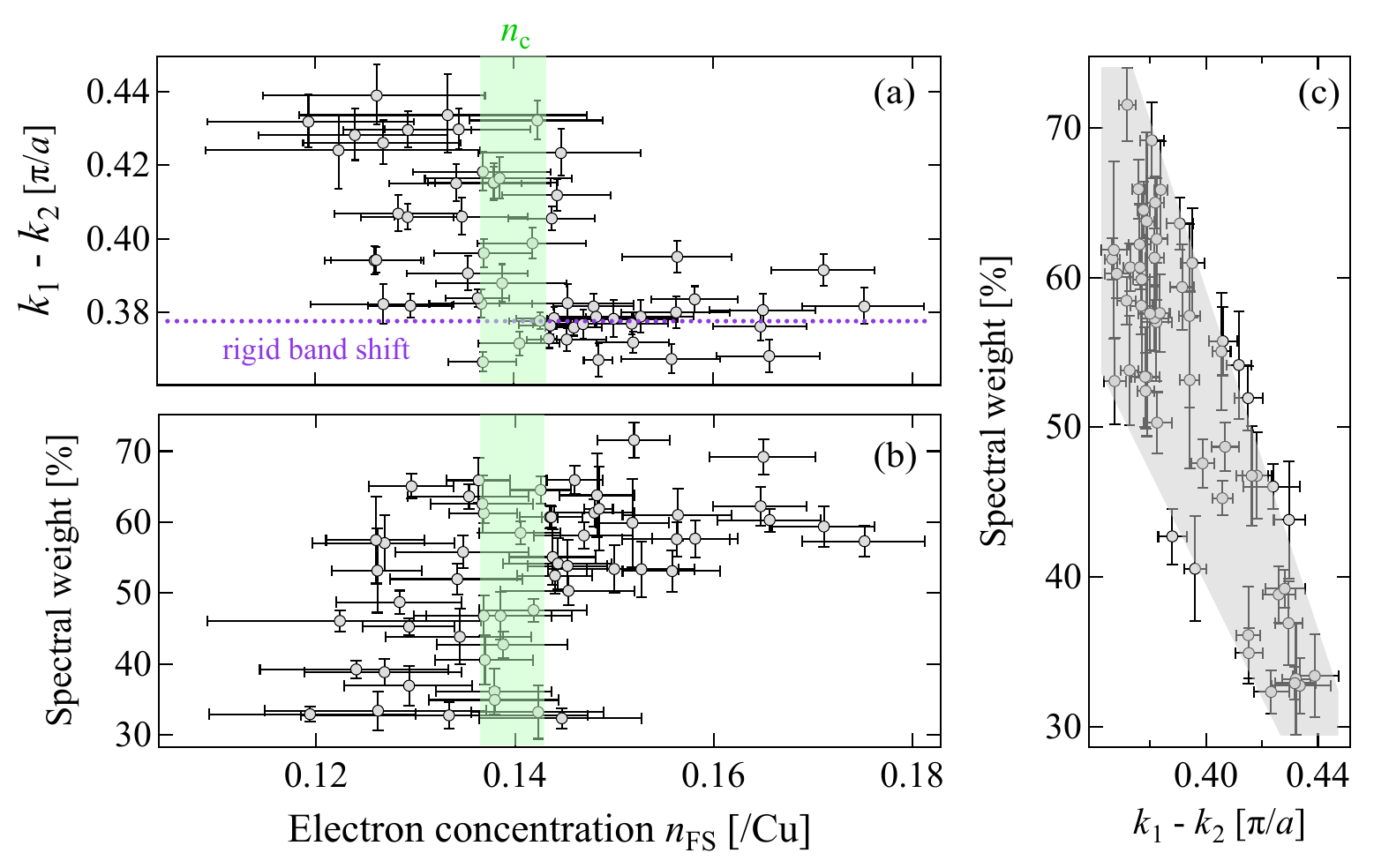}
  \end{center}
  \caption{\textbf{Correlation between the pseudogap and electron concentration.} 
	Electron-concentration dependence of (a) $k_{\rm 1}-k_{\rm 2}$ and 
	(b) spectral weight at the hot spot in panel (a), the case of the rigid band shift is 
	represented by the purple curve. Both $k_{\rm1}-k_{\rm 2}$ and spectral weight show 
	steep changes below $n_{\rm{FS}}\sim 0.14$ but appear saturated above $n_{\rm{FS}}\sim 0.14$. 
	(c) Spectral weight plotted against $k_{\rm 1}-k_{\rm 2}$. The two quantities have rather 
	monotonic negative correlation. 
	}
  \label{fig.4}
\end{figure*}

Figure~\ref*{fig.2} shows Fermi surface (FS) maps of the as-grown and protect-annealed PLCCO ($x=0.08$) 
samples at representative positions separated at least by 130 ${\rm\mu}$m among each other. 
For the as-grown sample [Fig.~\ref*{fig.2}(a)], spectral intensity is strongly suppressed at the hot spot 
(marked by circles), where the $\sqrt{2}\times\sqrt{2}$ antiferromagnetic Brillouin-zone boundary (AFM BZ) 
and Fermi surface cross. This indicates the opening of a pseudogap, and the Fermi surface 
can be approximated as an electron pocket centered around ($\pm\pi,0$) and ($0,\pm\pi$) 
\cite{armitage2001anomalous,armitage2003angle,matsui2007evolution}. 
The intensity suppression is equally observed irrespective of the sample 
position. In addition, the distance between the two antinodal segments (marked by dashed lines)
is virtually identical. In order to quantitatively evaluate the electron-doping level
from the FS maps around ($-\pi,0$), we phenomenologically employed the superellipse equation:
$\left|(k_x+\pi)/r\right|^{\alpha}+\left|k_y/r\right|^{\alpha}=1$, 
where $r$ and $\alpha$ are fitting parameters, which reproduces an electron pocket centered 
around ($-\pi,0$). All the electron-concentration values $n_{\rm{FS}}$ evaluated at eight 
different sample positions fall in the range of $0.073$-$0.095$, which is close to the nominal 
Ce concentration of $0.08$. On the other hand, for the annealed sample [Fig.~\ref*{fig.2}(b)], 
the intensity at the hot spot and the antinodal distance appear to vary depending on the sample 
positions, suggesting spatially inhomogeneous electronic states.\\ 
\indent The FSs of the annealed samples were measured at 56 different sample positions and fitted to the following 
two-dimensional tight-binding model as frequently performed for superconducting samples \cite{lin2021extended,horio2016suppression}:
\begin{equation}
	\begin{split}
		\epsilon - \mu =& \epsilon_0 - 2t(\cos k_xa+\cos k_ya) \\
	  &-4t'\cos k_xa\cos k_ya-2t''(\cos 2k_xa+\cos 2k_ya)
		\label{e2}
	\end{split}
\end{equation}
where \(t\), \(t'\), and \(t''\) are nearest-neighbor, second nearest-neighbor, 
and third nearest-neighbor hopping parameters, respectively. 
\(\epsilon_0\) represents the center of the band relative to the chemical potential \(\mu\). 
Throughout this fitting, the parameter $t''/t'$ was fixed at $-0.5$ as widely assumed for cuprates 
\cite{ikeda2009effects,horio2016suppression,horio2023influence}. In Fig.~\ref*{fig.3}(a), 
electron doping levels estimated from the fitted FS area, $n_{\rm FS}$, are plotted as color-coded 
points. The position of the points corresponds to the sample measurement position in the two-dimensional 
real space. The doping $n_{\rm{FS}}$ varies from $0.119$ to $0.175$ depending on the sample positions. 
These large values, which deviate from the nominal Ce concentration of $0.08$, can be understood if not 
only excess apical oxygen atoms but also oxygen atoms at the regular sites are removed 
by protect-annealing, effectively supplying electrons to the CuO$_2$ planes 
\cite{horio2016suppression,lin2021extended,horio2018angle,song2017electron,mang2004spin}. 
One might suspect that the sample surface degraded during the measurement, and thus the 
electron concentration varied as a function of time rather than of the sample position. However, 
such a possibility is excluded by the absence of a monotonic time-dependent trend in the 
estimated $n_{\rm FS}$ values (See Appendix). The significant spatial variation of the 
electron concentration suggests that oxygen diffusion by protect-annealing occurred in a 
spatially inhomogeneous fashion.\\
 In order to capture changes in the electronic structure beyond the rigid-band shift, we first turn to 
the anisotropic shift of Fermi wave number $k_{\rm F}$. Let us define nodal and antinodal $\vb{k}$ 
vectors $\vb{k_1}$ and $\vb{k_2}$, respectively, as shown in Fig.~\ref*{fig.3}(b). 
Since Fermi velocity is almost isotropic in electron-doped cuprates \cite{horio2020oxide,schmitt2008analysis}, 
the magnitude of the two vectors, $k_{\rm 1}$ and $k_{\rm 2}$ should change by an equal amount in 
the case of the rigid-band shift. Therefore, the difference between them, $k_{\rm 1}-k_{\rm 2}$, 
could be a parameter sensitive to non-trivial changes in the band structure. 
We determined $k_{\rm 1}$ and $k_{\rm 2}$ from the tight-binding fitted FS, 
and error bars of $k_{\rm 1}-k_{\rm 2}$ were estimated based on the standard deviation of 
the FS fitting. As shown in Fig.~\ref*{fig.4}(a), the value of 
$k_{\rm 1}-k_{\rm 2}$ varies by $\sim 0.06~\pi/a$ over the sample while typical error bars are 
$\sim0.005~\pi/a$, indicating the significance of the observed spatial variation. In addition, 
examining Figs.~\ref*{fig.3}(a) and (b) closely, the electron concentration and 
$k_{\rm 1}-k_{\rm 2}$ seem to be correlated; larger electron concentrations would yield 
smaller $k_{\rm 1}-k_{\rm 2}$.\\
 Another relevant quantity is spectral weight at the hot spot [Fig.~\ref*{fig.3}(c)]. 
We focused on an octant of the FS and determined the hot spot as 
a point that has the weakest spectral weight within $E_{\rm F}\pm$50 meV along the Fermi 
surface. Then, the intensity was normalized to that at $100$ meV in the same momentum cut. 
Again, the trend of the spatial variation [Fig.~\ref*{fig.3}(c)] resembles that of the 
electron concentration [Fig.~\ref*{fig.3}(a)], implying a correlation between the two quantities.\\
\indent Having evaluated the spatial variation of physical quantities, it is now possible to 
investigate their relationships by plotting one quantity versus another. The relation 
between $k_{\rm 1}-k_{\rm 2}$ and the electron concentration is shown in Fig.~\ref*{fig.4}(a). 
They are in general negatively correlated but with a kink at $n_{\rm FS}~\sim0.14~(=n_{\rm c})$. 
In $n_{\rm FS} > 0.14$, $k_{\rm 1}-k_{\rm 2}$ is rather constant and follows the calculated 
rigid-band behavior (shown as the purple dotted curve). In contrast, in $n_{\rm FS} < 0.14$, 
$k_{\rm 1}-k_{\rm 2}$ exhibits a sharp increase and deviates from the rigid-band curve, suggesting 
a dramatic FS transformation. On the other hand, the hot spot spectral weight is positively 
correlated with the electron concentration but again with a kink at $n_{\rm c}\sim 0.14$ 
[Fig.~\ref*{fig.4}(b)]. Plotting $k_{\rm 1}-k_{\rm 2}$ versus hot spot spectral weight, a negative 
and apparently monotonic correlation is found, implying a common physical origin underlying these two 
quantities [Fig.~\ref*{fig.4}(c)].

\section{Discussion}
\begin{figure}
  \begin{center}
    \includegraphics[width=0.47\textwidth]{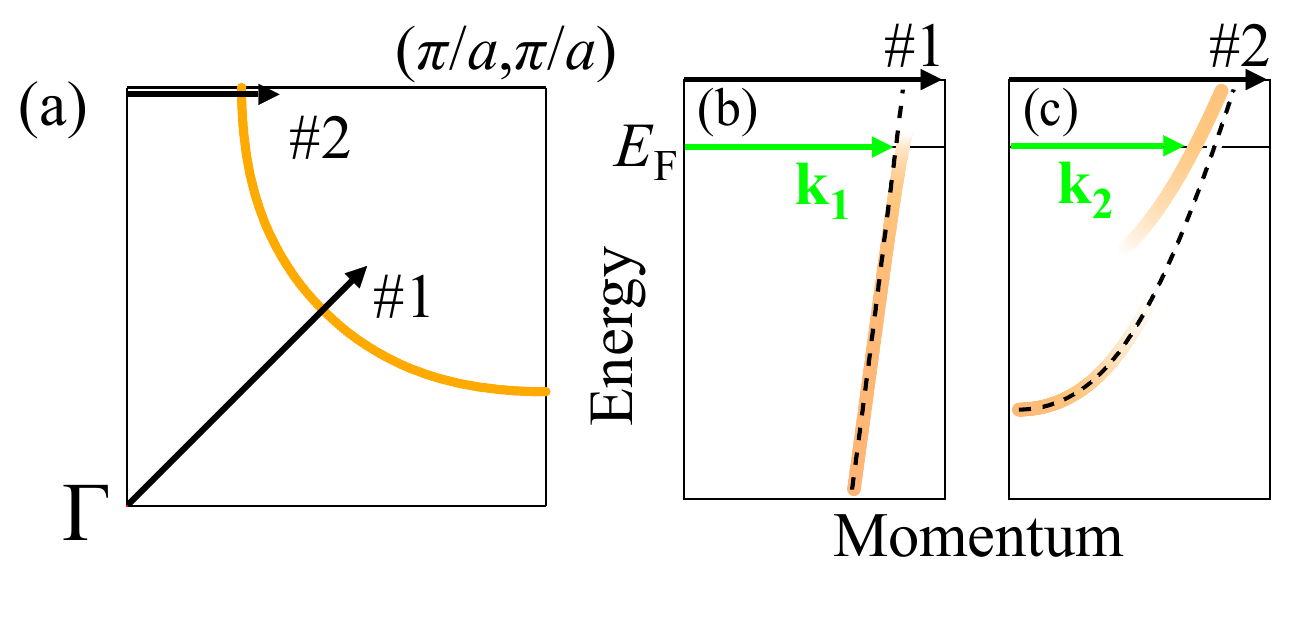}
  \end{center}
  \caption{	\textbf{Anisotropic shift of Fermi wave vectors by the pseudogap opening.} 
	(a) A schematic FS. (b),(c) Quasi-particle band dispersions along the nodal and antinodal 
	cuts [the black arrows in (a)], respectively. The dotted curves represent the case without 
	the pseudogap. The nodal and antinodal wave vectors are indicated by green arrows. When the 
	pseudogap opens, the nodal wave vector $\vb{k_1}$ is enlarged while the antinodal wave vector 
	$\vb{k_2}$ shrinks, resulting in the increase of $k_{\rm 1}-k_{\rm 2}$. 
	}
  \label{fig.5}
\end{figure}

\indent The reduction of the hot spot spectral weight with decreasing electron concentration can be 
readily understood as a result of enhanced pseudogap~\cite{matsui2007evolution,park2013interaction,song2017electron}. 
On the other hand, the unbalanced change of nodal and antinodal $k_{\rm F}$'s, captured 
by monitoring $k_{\rm 1}-k_{\rm 2}$, is not straightforward. The increase of $k_{\rm 1}-k_{\rm 2}$ 
suggests the decrease of the FS curvature and hence is translated into the decrease of $-t'/t$ in 
the framework of the tight-binding model [eq. \ref*{e2}]. Ikeda {\it et al.}~\cite{ikeda2009effects} 
pointed out a correlation between $-t'/t$ and antiferromagnetic (AFM) correlation in a series of electron-doped cuprates. 
By substituting rare-earth elements and thus applying a chemical pressure to cause lattice contraction in the 
in-plane direction, they observed the concomitant decrease of hot spot spectral weight and $-t'/t$. 
It was argued that the decrease of $-t'/t$ caused by the chemical pressure makes the Fermi 
surface more straight along the AFM BZ, and depletes the spectral weight at the hot spot through 
improved nesting. In the present case, the tendency of the hot spot spectral weight and FS 
curvature is consistent with the previous study~\cite{ikeda2009effects}. However, lattice 
parameter changes by annealing, particularly the in-plane ones, is in general negligibly small compared to that 
caused by replacing rare-earth elements \cite{radaelli1994evidence,schultz1996single,ikeda2009effects}, 
and would not significantly modify the hopping parameters. The observed changes in $k_{\rm 1}-k_{\rm 2}$ 
thus require other physical explanations.\\
 \indent The pseudogap opening can provide a natural explanation for the observed changes in $k_{\rm 1}-k_{\rm 2}$. 
Figure 5 schematically illustrates the opening of the pseudogap, whose momentum dependence is consistent with recent 
reports \cite{matsui2005angle,park2007electronic,ikeda2009effects,horio2018common}. Since the pseudogap opens above 
and below $E_{\rm F}$ at the node and the antinode, respectively, nodal $k_{\rm F}$ measured from ($0,0$) 
[$k_1$ in Fig.~\ref*{fig.5}(b)] becomes larger while the antinodal one measured from ($0,\pi$) [$k_2$ in Fig.~\ref*{fig.5}(c)] 
becomes smaller, resulting in the increase of $k_{\rm 1}-k_{\rm 2}$. The value $k_1-k_2$, therefore, should serve as a 
measure of the pseudogap magnitude. Negative and seemingly monotonic correlation between $k_{\rm 1}-k_{\rm 2}$ and the hot 
spot spectral weight [Fig.~\ref*{fig.4}(c)] is reasonable from this viewpoint. \\
\indent The plots in Figs. \ref*{fig.4} (a) and (b) suggest the existence of both of the following regions 
in a single sample after protect annealing: a moderately reduced region with a small electron concentration 
and enhanced pseudogap, and a strongly reduced region with a large electron concentration and 
suppressed pseudogap. This coexistence naturally explains the low superconducting fraction of the sample, 
though identifying the exact boundary between the superconducting and non-superconducting states requires better energy 
resolution to detect a superconducting gap of a few meV \cite{matsui2005angle,armitage2001superconducting,horio2019d,xu2023bogoliubov}. 
On the other hand, the kinks observed in the electron concentration dependence shown in Figs. \ref*{fig.4}(b) 
and (c) collectively imply a pseudogap critical point existing at $n_{\rm c}\sim 0.14$. 
Previously, abrupt changes in transport properties have been studied on annealed samples with varying Ce 
concentration $x$. For \PCCO (PCCO), a FS reconstruction was identified at $x_{\rm c}=0.165$ 
through Hall-effect \cite{dagan2004evidence} and thermopower measurements \cite{li2007evidence}, 
and ARPES studies on \NCCO ~\cite{matsui2007evolution,he2019fermi} revealed that the pseudogap is rapidly 
filled in around this doping. For \LCCO (LCCO), the critical doping is reduced to $x_{\rm c}=0.14$ according 
to Hall-effect \cite{sarkar2017fermi} and thermopower measurements \cite{mandal2019anomalous}. The 
difference in $x_{\rm c}$ likely originates from chemical pressure induced by substituting 
rare-earth elements. It is known that with larger ionic radius of the rare-earth element 
(smaller atomic number) superconductivity can be induced with less electron doping 
\cite{krockenberger2008superconductivity}. This is consistent with more rapid suppression of the 
pseudogap for LCCO than for PCCO \cite{zimmers2005infrared,tang2021suppression}. In the present 
case, variation of the electron concentration was realized through the removal of oxygen atoms. 
Nevertheless, the present value of $n_{\rm c}\sim 0.14$ for PLCCO is close to the critical Ce 
concentration value for LCCO ($x_{\rm c}=0.14$) and smaller than that for PCCO ($x_{\rm c}=0.165$). 
This suggests that, once sufficiently annealed, the pseudogap property does not strongly depend on the exact 
mechanism of electron doping, whether by Ce substitutions or oxygen removal \cite{horio2018angle}. 
It is of note that such a critical behavior is identified from the measurement of the single 
sample by utilizing the large amount of the dataset obtained from spatially varying electronic states.\\
\indent However, one also needs to notice that Figs. \ref*{fig.4} (a) and (b) do not necessarily 
reflect the genuine electron-doping dependence of the pseudogap. Since the $n_{\rm FS}$ of the 
as-grown sample is not smaller than the nominal Ce concentration, excess oxygen atoms at the apical site 
apparently do not affect the carrier concentrations \cite{tsukada2005role}. Therefore, oxygen defects 
created in the (Pr,La,Ce)$_2$O$_2$ or CuO$_2$ layers by annealing should be responsible for the variation 
of the electron concentration. The horizontal axis of Figs.~\ref*{fig.4} (a) and (b) is thus equivalent to 
the amount of the defects at the regular sites. While the amount of oxygen atoms removed from the apical site 
and regular sites should be positively correlated, the correlation would not be necessarily perfect. 
In fact, the values on the vertical axis of Figs.~\ref*{fig.4} (a) and (b), which represent pseudogap magnitude, 
are widely distributed even at a fixed $n_{\rm FS}$ in the region of $n_{\rm FS} < 0.14$. It is possible that 
this distribution arises from a subtle difference in the amount of residual apical oxygen atoms. Disorder of 
electrostatic potential at the apical site causes electron localization \cite{adachi2013evolution}, which could 
enhance the pseudogap by reinforcing the effect of electron correlation \cite{horio2018common}. The present 
results on the pseudogap may thus contain significant influence from apical oxygen atoms while the effect becomes 
difficult to pursue when the oxygen reduction progresses further to increase the electron concentration beyond $0.14$. 

\section{Conclusion}
 In summary, we investigated the influence of protect-annealing on the electronic structure of 
Ce-underdoped cuprate PLCCO ($x=0.08$) with small superconducting volume. By performing space-resolved 
ARPES measurements using a microbeam, it was found that both the electron concentration $n_{\rm FS}$ 
and the pseudogap magnitude are spatially varying probably due to the inhomogeneous diffusion of oxygen atoms. 
ARPES spectra collected at 56 different sample positions were utilized to find a correlation between $n_{\rm FS}$ 
and the pseudogap magnitude. The pseudogap critical point of $n_{\rm c}\sim 0.14$ is compatible with expectations 
from Ce doping dependences previously studied on LCCO and PCCO, suggesting the major influence of electron 
doping on the electronic structure. Still, the effect of apical oxygen atoms on the pseudogap also seems 
significant particularly at low electron concentrations.

\section*{Acknowledgements}
The authors wish to thank A. Fujimori, M. Fujita, T. Taniguchi, and S. Ideta for helpful discussions. 
This work was supported by JSPS KAKENHI Grant No. JP24KK0061, No. JP21K13872 and No. JP19H01841. 
We acknowledge MAX IV Laboratory for time on Beamline BLOCH under Proposals No. 20220314 and 
20230779. Research conducted at MAX IV, a Swedish national user facility, is supported by the 
Swedish Research council under Contract No. 2018-07152, the Swedish Governmental Agency for 
Innovation Systems under contract No. 2018-04969, and Formas under Contract No. 2019-02496.\\

\appendix
	\section*{Appendix: Influence of surface degradation on the evaluation of inhomogeneity}
\begin{figure}
		\begin{center}
			\includegraphics[width=0.47\textwidth]{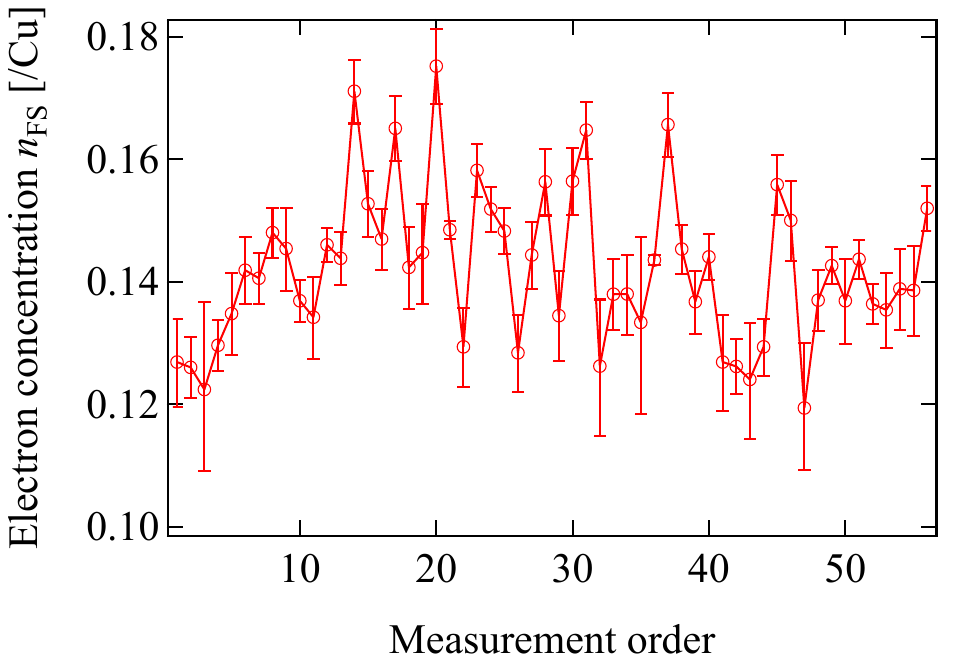}
		\end{center}
		\caption{\textbf{Estimated electron concentrations plotted in order of measurements.} 
		There is no monotonic trend and thus the possibility of surface degradation is 
		excluded as the primary origin of the observed position dependence.
		}
		\label{appfig}
	\end{figure}

	Figure \ref*{appfig} shows the electron concentration $n_{\rm FS}$ per Cu atom estimated for 
	the protect-annealed sample at 56 different positions in a time series. There is no monotonic 
	trend in $n_{\rm FS}$, which ensures that the observed position dependence is not due to 
	surface degradation that should progress monotonically with time.\\

\end{document}